\documentclass{ws-p8-50x6-00}

\def\mbox{\hbox}           

\def\deg{\ifmmode ^\circ                
         \else $^\circ$
         \fi
         \hskip -0.1truecm}
\def\degd#1.#2{                         
               \ifmmode {#1^{\hskip 0.05em\circ}\hskip-0.42em.\hskip0.08em#2}
               \else {#1$^{\hskip 0.05em\circ}\hskip-0.42em.\hskip0.08em$#2}
               \fi
              }
\def\mind#1.#2{                         
               \ifmmode {#1^{\hskip 0.05em\prime}\hskip-0.35em.\hskip0.05em#2}
               \else {#1$^{\hskip 0.05em\prime}\hskip-0.35em.\hskip0.05em$#2}
               \fi
              }
\def\secd#1.#2{                         
               \ifmmode {#1^{\prime\prime}\hskip-0.46em.\hskip0.12em#2}
               \else {#1$^{\prime\prime}\hskip-0.46em.\hskip0.12em$#2}
               \fi
              }
\def\timsecd#1.#2{                      
                  \ifmmode {#1^{\rm s}\hskip-0.39em.\hskip0.08em#2}
                  \else {$#1^{\rm s}\hskip-0.39em.\hskip0.08em#2$}
                  \fi
                 }
\def\hms#1h#2m#3s{                      
                  \relax
                  \ifmmode #1^{\rm h}\,#2^{\rm m}\,#3^{\rm s}
                  \else \hbox{$#1^{\rm h}\,#2^{\rm m}\,#3^{\rm s}$}
                  \fi
                 }
\def\dms#1d#2m#3s{                      
                  \relax
                  \ifmmode #1^\circ\,#2^{\prime}\,#3^{\prime\prime}
                  \else \hbox{$#1^\circ\,#2^{\prime}\,#3^{\prime\prime}$}
                  \fi
                 }
\def\dmsd#1d#2m#3.#4s{                  
                      \relax
                      \ifmmode #1^\circ\,#2^{\prime}\,#3^{\prime\prime}
                               \hskip-0.46em.\hskip0.12em#4
                      \else \hbox{$#1^\circ\,#2^{\prime}\,#3^{\prime\prime}
                            \hskip-0.46em.\hskip0.12em#4$}
                      \fi
                     }
\def\hm#1h#2m{                          
              \relax
              \ifmmode #1^{rm h}\,#2^{\rm m}
              \else \hbox{$#1^{\rm h}\,#2^{\rm m}$}
              \fi
             }
\def\dm#1d#2m{                          
              \relax
              \ifmmode #1^\circ\,#2^{\prime}
              \else \hbox{$#1^\circ\,#2^{\prime}$}
              \fi
             }
\def\hmsd#1h#2m#3.#4s{                  
                      \relax
                      \ifmmode #1^{\rm h}\,#2^{\rm m}\,#3^{\rm s}
                               \hskip-0.39em.\hskip0.08em#4
                      \else \hbox{$#1^{\rm h}\,#2^{\rm m}\,#3^{\rm s}
                            \hskip-0.39em.\hskip0.08em#4$}
                      \fi
                     }
\def\hmd#1h#2.#3m{                  
                  \relax
                  \ifmmode #1^{\rm h}\,#2^{\rm m}
                           \hskip-0.55em.\hskip0.22em#3
                  \else \hbox{$#1^{\rm h}\,#2^{\rm m}
                        \hskip-0.55em.\hskip0.22em#3$}
                  \fi
                 }
\def\mg{\relax                          
        \ifmmode ^{\rm m}
        \else $^{\rm m}$
        \fi
       }
\def\mgd#1.#2{                          
              \relax
              \ifmmode #1^{\rm m}
                       \hskip-0.55em.\hskip0.22em#2
              \else \hbox{#1$^{\rm m}
                    \hskip-0.55em.\hskip0.22em$#2}
              \fi
             }

\def\la{\mathrel{\hbox{\rlap{\hbox{\lower4pt\hbox{$\sim$}}}\hbox{$<$}}}}
\def\ga{\mathrel{\hbox{\rlap{\hbox{\lower4pt\hbox{$\sim$}}}\hbox{$>$}}}}

\def\unitspace{\;}                      

\def\un#1{\ifmmode \unitspace\mbox{\rm #1} 
          \else $\unitspace$#1
          \fi}
\def\pun#1#2{\ifmmode \unitspace\mbox{\rm #1}^{#2} 
             \else $\unitspace$#1$^{#2}$
             \fi}

\def\kms{\un{km}\pun{s}{-1}}          
\def\Lsun{\ifmmode \un{L}_{\odot}     
          \else $\un{L}_{\odot}$
          \fi}
\def\Msun{\ifmmode \un{M}_{\odot}     
          \else $\un{M}_{\odot}$
          \fi}
\def\mum{\ifmmode \unitspace\mu\mbox{\rm m} 
         \else $\unitspace\mu$m
         \fi}
\def\sqarcsec{\ifmmode \unitspace\Box''    
              \else $\unitspace\Box''$     
              \fi} 

\def\Bp{\relax                            
        \ifmmode B_{||}                   
        \else $B_{||}$
        \fi}
\def\Bt{\relax                            
        \ifmmode B\!_{\perp}              
        \else $B\!_{\perp}$               
        \fi}
\def\Gcr{\relax                           
         \ifmmode \Gamma\!_{\rm cr}       
         \else $\Gamma\!_{\rm cr}$
         \fi}
\def\ICII{\relax                          
          \ifmmode I_{[\CII]}             
          \else $I_{[\CII]}$
          \fi}
\def\LHtwo{\relax                                 
           \ifmmode L_{\mbox{\rm\scriptsize H}_2} 
           \else $L_{\mbox{\rm\scriptsize H}_2}$  
           \fi}
\def\LLya{\relax                          
          \ifmmode L_{{\rm Ly}\,\alpha}   
          \else $L_{{\rm Ly}\,\alpha}$
          \fi}
\def\MHtwo{\relax                                 
           \ifmmode M_{\mbox{\rm\scriptsize H}_2} 
           \else $M_{\mbox{\rm\scriptsize H}_2}$  
           \fi}
\def\MHtwodot{\relax                                       
              \ifmmode \dot{M}_{\mbox{\rm\scriptsize H}_2} 
              \else $\dot{M}_{\mbox{\rm\scriptsize H}_2}$  
              \fi}                                         
\def\Mstardot{\relax                      
              \ifmmode \dot{M}_{\ast}     
              \else $\dot{M}_{\ast}$      
              \fi}
\def\nHI{\relax                                      
         \ifmmode n_{\mbox{\scriptsize\rm H\,\sc I}} 
         \else $n_{\mbox{\scriptsize\rm H\,\sc I}}$
         \fi}
\def\nHtwo{\relax                                
           \ifmmode n_{{\mbox{\scriptsize H}}_2} 
           \else $n_{{\mbox{\scriptsize H}}_2}$  
           \fi}
\def\rhostardot{\relax                         
                \ifmmode \dot{\rho}_{\ast}     
                \else $\dot{\rho}_{\ast}$      
                \fi}
\def\rhoZdot{\relax                          
             \ifmmode \dot{\rho}_{\rm Z}     
             \else $\dot{\rho}_{\rm Z}$      
             \fi}

\def\sou#1#2{\relax                       
             \ifmmode {\rm #1}\,{\rm #2}  
             \else #1$\,$#2
             \fi}

\def\qu#1#2{\relax                          
            \ifmmode #1_{\rm #2}            
            \else $#1_{\rm #2}$
            \fi}

\def\CO#1{\ifnum#1=0                    
           \ifmmode \mbox{\rm CO}
           \else {\rm CO}
           \fi
          \else
           \ifnum#1<15
            \ifmmode ^{#1}\mbox{\rm CO}
            \else $^{#1}${\rm CO}
            \fi
           \else
            \ifmmode \mbox{\rm C}^{#1}\mbox{\rm O}
            \else {\rm C}$^{#1}${\rm O}
            \fi
           \fi
          \fi}

\def\COp{\ifmmode \mbox{\rm CO}^+           
         \else {\rm CO}$^+$                 
         \fi}

\def\CS#1{\ifnum#1=0                    
           \ifmmode \mbox{\rm CS}
           \else {\rm CS}
           \fi
          \else
           \ifnum#1<15
            \ifmmode ^{#1}\mbox{\rm CS}
            \else $^{#1}${\rm CS}
            \fi
           \else
            \ifmmode \mbox{\rm C}^{#1}\mbox{\rm S}
            \else {\rm C}$^{#1}${\rm S}
            \fi
           \fi
          \fi}

\def\HCOp{\ifmmode \mbox{\rm HCO}^+          
          \else {\rm HCO}$^+$                
          \fi}
\def\Hthreep{\ifmmode \mbox{\rm H}_3^+         
             \else {\rm H}$_3^+$               
             \fi}

\def\Htwo{\ifmmode \mbox{\rm H}_2              
          \else {\rm H}$_2$                    
          \fi}

\def\HtwoO{\ifmmode \mbox{\rm H}_2\mbox{\rm O} 
           \else {\rm H}$_2${\rm O}            
           \fi}

\def\ion#1#2{\ifmmode \mbox{{\rm #1}}\,\mbox{{\sc #2}} 
        \else {\rm #1}$\,${\sc #2}
        \fi}

\def\rec#1#2{\if#2a                            
              \ifmmode \mbox{{\rm #1}}\alpha   
              \else {\rm #1}$\alpha$
              \fi
             \fi
             \if#2b
              \ifmmode \mbox{{\rm #1}}\beta
              \else {\rm #1}$\beta$
              \fi
             \fi
             \if#2g
              \ifmmode \mbox{{\rm #1}}\gamma
              \else {\rm #1}$\gamma$
              \fi
             \fi}

\newcommand{\tabref}[1]{Table~\protect\ref{#1}}
\newcommand{\figref}[1]{Fig.~\protect\ref{#1}}

\newcommand{\eqref}[1]{Eq.~$\left(\protect\ref{#1}\right)$}

\begin{document}

\title{Can dusty Lyman break galaxies produce the
submillimeter counts and background?\\
Lessons from lensed Lyman break galaxies}

\author{Paul P.~van der Werf, Kirsten Kraiberg Knudsen, Ivo Labb\'e
and Marijn Franx}

\address{Leiden Observatory, P.O.~Box~9513, NL~-~2300~RA~Leiden, 
The Netherlands\\E-mail: pvdwerf,kraiberg,ivo,franx@strw.leidenuniv.nl}


\maketitle

\abstracts{
Can the submillimeter counts and background be produced by applying a
locally derived extinction correction to the population of Lyman break
galaxies? We investigate the submillimeter emission of two strongly
lensed Lyman break galaxies ($\sou{MS}{1512{+}36}$-cB58 and
$\sou{MS}{1358{+}62}$-G1) and find that the procedure that is used to
predict the submillimeter emission of the Lyman break galaxy population
overpredicts the observed $850\mum$ fluxes by up to a factor of 14. 
This result calls for caution in applying
local correlations to distant galaxies. It also shows that large
extinction corrections on Lyman break galaxies should be viewed with
skepticism. It is concluded that the Lyman break galaxies may
contribute to the submillimeter background at the 25 to 50\% level.
The brighter submillimeter galaxies making up the rest of the
background are either not detected in optical surveys, or if they are
detected, their submillimeter emission cannot be reliably estimated
from their rest-frame ultraviolet properties.
}

\section{Submillimeter emission from Lyman break galaxies}
\label{sec.LBGs}

Measurements of the cosmic star formation rate density (SFRD) based on
surveys for Lyman break galaxies (LBGs) are affected by 
extinction, and attempts to
correct for this effect lead to a substantial upwards revision of the 
SFRD\null.~\cite{Meureretal99} 
Since the light absorbed in the ultraviolet (UV) is reradiated in the 
far-infrared (FIR), LBGs affected by extinction
must emit FIR radiation, which will contribute to the submillimeter
background.

The expected submillimeter emission from the population of LBGs has
recently been estimated based on the observed (if not entirely
understood) correlation between the spectral index $\beta$ in the
UV (defined by the relation
$f_\lambda\propto\lambda^\beta$) and the ratio of FIR to UV flux in a
sample of local galaxies.~\cite{Meureretal99} Applying this relation to the LBG
population, Adelberger \& Steidel~\cite{AdelbergerSteidel00} 
found that the submillimeter counts and integrated 
background can be accounted for. As noted by these authors, these
estimates are uncertain, since 
the validity of the $\beta$-FIR/UV correlation
at high redshift has not been established, and 
the distribution of $\beta$ values and
its dependence on magnitude, and the luminosity function at faint
magnitudes are poorly constrained. 
It is therefore necessary to verify this analysis by
direct submillimeter observations of LBGs.

\begin{table}[t]
\caption{Predicted (from UV color and magnitude) and observed
submillimeter emission of lensed LBGs. Luminosities (not corrected for
gravitational amplification) have been derived
for $H_0=50\kms\pun{Mpc}{-1}$ and $q_0=0.5$. The other results do not
depend on cosmology. Upper limits represent $3\sigma$.\label{tab.results}}
\begin{center}
\footnotesize
\begin{tabular}{|c|c|c|}
\hline
{} &\raisebox{0pt}[13pt][7pt]{$\sou{MS}{1512{+}36}$-cB58} &
\raisebox{0pt}[13pt][7pt]{$\sou{MS}{1358{+}62}$-G1}\\
\hline
& & \\[-5pt]
amplification factor & $\sim50$ & $\sim10$ \\
$\beta$ (observed) & $-0.74\pm0.1$ & $-1.63\pm0.1$ \\
$A_{1600}$ (predicted) & $3.0\pm0.2$ & $1.2\pm0.2$ \\
$\qu{S}{FIR}/\qu{S}{1600}$ (predicted) & 18 & 2.4 \\
$\qu{L}{FIR}$ (predicted) & $3.9\cdot10^{13}\Lsun$ &
$3.3\cdot10^{12}\Lsun$ \\
$S_{850}$ (predicted) & $58\un{mJy}$ & $5\un{mJy}$ \\
$S_{850}$ (observed)  & $4.2\pm0.9\un{mJy}$ & $<4\un{mJy}$ \\
$\qu{L}{FIR}$ (derived) & $2.8\cdot10^{12}\Lsun$ &
$<2.6\cdot10^{12}\Lsun$ \\
$\qu{S}{FIR}/\qu{S}{1600}$ (derived) & 1.3 & $<1.9$ \\
$A_{1600}$ (derived) & 0.8 & $<1.0$ \\[5pt]
\hline
\end{tabular}
\end{center}
\end{table}

\begin{figure}[t]
\epsfxsize=20pc 
\begin{center}
\epsfbox{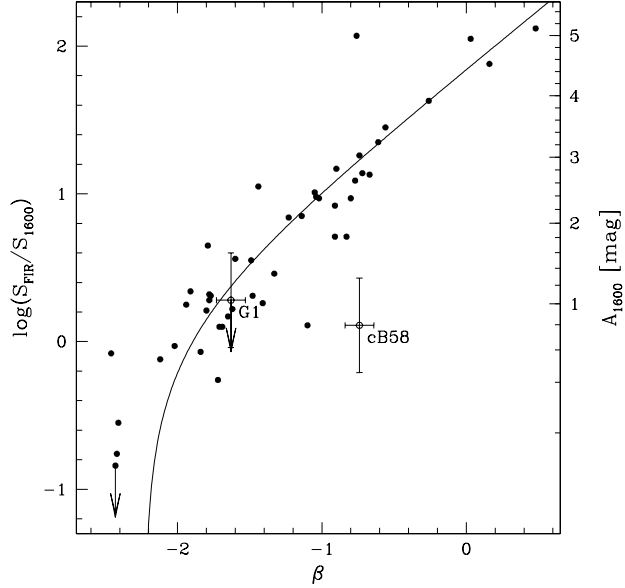} 
\end{center}
\caption{The relation between $\qu{S}{FIR}/\qu{S}{1600}$ and $\beta$
for local UV-selected galaxies (filled circles) with the
best-fitting parametrization,~\protect\cite{Meureretal99} and the positions of
cB58 and G1 with respect to this relation. \label{fig.beta}}
\end{figure}

\section{Lessons from strongly lensed Lyman break galaxies}
\label{sec.results}

The predicted $850\mum$ fluxes for most LBGs based on their UV
properties~\cite{AdelbergerSteidel00} are $1\un{mJy}$ or less,
which is too faint for current instrumentation.
Submillimeter observations of individual LBGs are significantly
easier if strongly lensed LBGs are targeted. We have used SCUBA on the
JCMT to observe two strongly lensed LBGs: the object cB58 at $z=2.72$ lensed by
the cluster $\sou{MS}{1512{+}36}$~\cite{Yeeetal96} and the
object G1 at $z=4.92$ lensed by the cluster 
$\sou{MS}{1358{+}62}$.~\cite{Franxetal97} Rest-frame UV colors
indicate significant reddening in both of these 
objects.~\cite{Ellingsonetal96,Soiferetal98}
For both galaxies, the method of predicting the FIR emission
based on the UV properties 
$\beta$~\cite{Pettinietal00,Soiferetal98} implies
strong $850\mum$ emission (\tabref{tab.results}).
The results of our SCUBA measurements are given in
\tabref{tab.results}. The object cB58 is detected at the $4.7\sigma$
level; the object G1 was not detected. 
In both cases the procedure of predicting the
submillimeter flux from the observed color and magnitude in the
rest-frame UV~\cite{AdelbergerSteidel00} overpredicts the
submillimeter emission. The
magnitude of the discrepancies is illustrated in \figref{fig.beta}.
For G1 the discrepancy is not significant,
given the scatter in the $\beta$-FIR/UV relation, but the factor 14
discrepancy for cB58 is highly significant. 

Since this discrepancy is so large, 
it cannot be attributed to
observational uncertainties. However, an element of uncertainty
in the analysis is introduced by differential lensing, if the
effective amplification factor of cB58 in the UV is a factor 10 
higher than that
in the FIR\null. A large discrepancy can only be introduced 
if most of the UV emission comes from the most strongly amplified
portions of the source near the caustic, while most of the FIR
emission comes from more weakly lensed regions. This situation would
require a very different distribution of FIR and UV emission. However, an
extinction correction based on UV properties of one region of a galaxy
will not be able to predict the submillimeter emission in a completely 
different
region of the system. Therefore, if differential lensing plays a major
role, the physical basis of using the $\beta$-FIR/UV correlation disappears.
For cB58, a lensing model based on new HST data shows
that the amplification factor is at least a factor of 5 at every
position, and that the intensity-weighted amplification factor in the UV is
about a factor of 25 (on both sides of the fold arc, so that the total
amplification is approximately a factor 50). 
Thus differential lensing can account for a
discrepancy of at most a factor of 5, but probably much less, since UV
and FIR emission should have a similar morphology for the
$\beta$-FIR/UV correlation to work.

\section{Discussion and conclusions}

These results demonstrate that attempts to produce the submillimeter counts and
background based on an extinction correction applied to the LBG 
population~\cite{AdelbergerSteidel00} are fraught with considerable
uncertainty, and fail in the case of the two lensed LBGs discussed
here. While our sample is small, the
results argue against the validity of the low redshift $\beta$-FIR/UV
correlation in the case of LBGs. 
Even if the local correlation were valid for high redshift
galaxies as well, it still would not be able to produce the brighter
submillimeter galaxies. These objects have luminosities
putting them in the class of the ultraluminous infrared galaxies
(ULIGs), which do not follow the $\beta$-FIR/UV correlation, as shown
by recent HST-STIS data of a sample of nearby ULIGs. Since these
galaxies already account for $\sim50\%$ of the submillimeter
background, it is not possible for the LBGs to produce a dominant
fraction (let alone all) of the submillimeter background. A more
likely situation is that the LBGs account for 25 to 50\% of the
submillimeter background as indicated by the faint structure in the
HDF at $850\mum$,~\cite{Peacocketal00} but that the dominant part of
the submillimeter background is made by a small number of ULIGs, which
are either not detected in LBG surveys, or if they are detected,
cannot be reliably corrected for extinction, in the same way that
local ULIGs cannot be extinction corrected based on UV data.

In summary therefore, these results support the view that 
the brighter $850\mum$ galaxies making at least
50\% of the submillimeter background, form a population 
which cannot be reliably reproduced by extinction corrections applied
to the LBG population.

\end{document}